\documentstyle[eqsecnum,aps]{revtex}

\begin{document}
\title{Crucial Dependence of ``Precarious'' and ``Autonomous'' $\phi^4$s \\Upon
       the Normal-ordering Mass}
\author{Wen-Fa Lu }
\address{Department of Applied Physics, Shanghai Jiao Tong University,
         Shanghai 200030, China
         \thanks{E-mail: wenfalu@online.sh.cn} }
\maketitle

\begin{abstract}
Using the Gaussian wave-functional approach with the normal-ordering
renormalization prescription, we show that for the (3+1)-dimensional massive
$\lambda\phi^4$ theory, ``precarious'' and ``autonomous'' $\phi^4$s can exist
if and only if the normal-ordering mass is equal to the classical masses at the
symmetric and asymmetric vacua, respectively.
\end{abstract}
\vspace{24pt}

The triviality problem of quantum field theories, for example, the
$\lambda\phi^4$ theories, is very important and has been receiving lots of
investigations since the early 1970's \cite{1}. In general, it is believed
that $(D+1)$-dimensional field theories $(D\ge 3)$ have been proved to be
trivial ($i.e$, to be non-interacting or else inconsistent). In the 1980's,
nevertheless, P. M. Stevenson and his collaborators took a new view of the
(3+1)-dimensional $\lambda\phi^4$ theory and accordingly proposed two
non-trivial $\lambda\phi^4$ theories : ``percarious $\phi^4$'' and ``autonomous
$\phi^4$''. Within the framework of the Gaussian wave-functional approch (GWFA),
for the case of the negative and infinitesimal bare coupling, Stevenson
renormalized the (3+1)-dimensional $\lambda\phi^4$ theory by a set of
renormalization conditions of the mass and coupling parameters at the zero
vacuum expectation value (VEV) of field operator, and obtained a non-trivial
theory called ``precarious $\phi^4$'' \cite{2}; for the case of the positive
and infinitesimal bare coupling, Stevenson and Tarrach also renormalized the
(3+1)-dimensional $\lambda\phi^4$ theory by another set of renormalization
conditions of the mass and coupling parameters at the zero VEV of field
operator and further by an infinite wavefunction renormalization condition, and
obtained the other non-trivial theory called ``autonomous $\phi^4$'' \cite{3}.
``Precarious $\phi^4$'' has a stable, symmetric phase \cite{2}, whereas
``autonomous'' $\phi^4$ can exhibit either a massless symmetry phase or a
massive broken-symmetry phase \cite{3}. Since the two non-trivial theories were
proposed, they have attracted much attention. These two non-trivial theories
were demonstrated to exist in O(N)-symmetric $\lambda\phi^4$ theory \cite{4},
and their generalizations to many other complicated models were made, too
\cite{5}. Furthermore, it was shown that in the context of dimensional
regularization, ``precarious $\phi^4$'' arises as the (D+1)$\to (4)_+$ limit of
the (D+1)-dimensional $\lambda \phi^4$ theory with $D>3$, whereas ``autonomous
$\phi^4$'' can be understood as the (D+1)$\to (4)_-$ limit of the
(D+1)-dimensional $\lambda \phi^4$ theory with $D<3$ \cite{6}. Again,
``precarious'' and ``autonomous'' $\phi^4$s were discussed by some methods
beyond the Gaussian approximation, and the existence of ``autonomous $\phi^4$''
was still demonstrated but those methods did not all confirm the appearance of
``precarious $\phi^4$'' \cite{7}. Besides, ``autonomous $\phi^4$'' arising from
the classical pure $\lambda\phi^4$ theory (massless) was believed to not
conflict with ``Triviality'', and used in relevant Models to predict a 2.2 Tev
Higgs boson \cite{8}.

To our knowledge, no work discussed the dependences of ``precarious'' and
``autonomous'' $\phi^4$s upon renormalization points. This is perhaps
because results obtained from the GWFA with the standard renormalization
prescription \footnote{In the standard renormalization prescription of the
GWFA, the definitions of renormalized physical parameters resemble those in the
classical theory concerned.} \cite{9} are exactly renormalization-group
invariant \cite{2} (1985, Sect.III). However, although ``precarious $\phi^4$''
was obtained according to the standard renomalization prescription, an
approximation was made in renormalizing the coupling parameter \cite{2}, and as
for the ``autonomous $\phi^4$'', an unusual set of renormalization conditions
was proposed within the framework of the GWFA \cite{3}. On the other hand, as
was mensioned above, it was at a peculiar renormalization point, the zero VEV
of field operator, that both ``precarious'' and ``autonomous'' $\phi^4$s were
demonstrated to exist \cite{2,3}. Therefore, it is necessary and important to
consider whether ``precarious'' and ``autonomous'' $\phi^4$s would exist or not
when the (3+1)-dimensional $\lambda \phi^4$ theory is renormalized at any other
renormalization points. In this letter, we intend to address this problem.

For the above problem, the renormalization prescriptions utilized by Stevenson
$et \ al$ \cite{2,3} are not too convenient because it must be executed at a
explicit renormalization point given previously. Nevertheless, within the
framework of the GWFA, executing the standard renormalization prescription
\cite{9} is equivalent to executing Coleman's renormalization prescription
\cite{10,11,12,13}, and further, executing a renormalization procedure at
different renomalized points amounts to choosing different values of the
normal-ordering mass in Coleman's normal-ordering renormalization prescription
\cite{11}. Thus, if Coleman's normal-ordering renormalization prescription is
adopted, the dependence of a field theory on renomalization points will be
equivalent to the dependence of that theory on the normal-ordering mass.
Luckily, Coleman's renormalization prescription can be performed at any
normal-ordering mass which needn't be explicitly given, and this is evidently
convenient for the purpose here. Hence, in order to know if the existences of
``precarious'' and ``autonomous'' $\phi^4$s depend upon renormalization points,
it is sufficient to renormalize the (3+1)-dimensional $\lambda\phi^4$ theory on
the basis of Coleman's normal-ordering renormalization prescription and pay
attention to the dependence of ``precarious'' and ``autonomous'' $\phi^4$s upon
values of the normal-ordering mass instead of the renormalization points.

In the following, based on Coleman's normal-ordering renormalization
prescription, we shall analyse the Gaussian effective potential (GEP) of the
(3+1)-dimensional massive $\lambda \phi^4$ theory, manage to turn the GEP
finite, and accordingly show that ``precarious'' and ``autonomous'' $\phi^4$s
appear only at two particular values of the normal-ordering mass,respectively.

Consider the massive single-component $\lambda\phi^4$ field thory in (3+1)
dimensions \cite{2,3},
\begin{equation}
{\cal L}={\frac {1}{2}}\partial_\mu \phi_x \partial^\mu \phi_x-
                      {\frac {1}{2}}m^2\phi_x^2 - \lambda\phi_x^4 \; ,
\end{equation}
where $\phi_x\equiv\phi(\vec{x})$ and $\vec{x}$ is the position vector in the
three-dimensional space. In the fixed-time functional Schr\"odinger picture,
the Hamiltonian reads
\begin{equation}
H=\int_x {\cal H}_x=\int_x \{{\frac {1}{2}}\Pi^{2}_x
   +{\frac {1}{2}}(\nabla\phi_x)^2
   +{\frac {1}{2}}m^2\phi_x^2 + \lambda\phi_x^4\} \;,
\end{equation}
where, $\int_x\equiv\int d^3\vec{x}$, $\nabla$ is the gradient operator and
$\Pi_x\equiv -i{\frac {\delta}{\delta \phi_x}}$ conjugate to the field operator
$\phi_x$. The mass and coupling parameters $m$ and $\lambda$ govern the
classical physics of the system. When $m^2>0$ the classical vacuum is symmetric
and the classical mass squared is $m^2$, but when $m^2<0$ the classical vacuum
is spontaneously symmetry-breaking and the classical mass squared is $-2m^2$.

Take as an ansatz the general Gaussian wave-functional \cite{9}
\begin{equation}
|\varphi> \to \Psi[\phi;\varphi,{\cal P},f]= N_f exp\{i\int {\cal P}_x\phi_x
         -{\frac {1}{2}}\int_{x,y}(\phi_x - \varphi_x)f_{xy}
         (\phi_y - \varphi_y)\} \;,
\end{equation}
where ${\cal P}_x, \varphi_x$ and $f_{xy}$ are variational
parameter-functions. $N_f$ is some normalization constant, and depends upon
$f_{xy}$. With respect to any normal-ordering mass $M$ (an arbitrary positive
constant with the dimension of mass), one can Nomal-order the Hamiltonian
density in Eq.(2) according to Ref.~\cite{10}, and has
\begin{eqnarray}
{\cal N}_M[{\cal H}_x]={\cal H}_x -
  3\lambda I_1(M^2)\phi^2_x+
  {\frac {3}{4}}\lambda I^2_1(M^2) -{\frac {1}{4}}m^2I_1(M^2)
  -{\frac {1}{2}}I_0(M^2) + {\frac {1}{4}}M^2I_1(M^2)
  \;
\end{eqnarray}
with the notation
\begin{equation}
I_n(M^2)=\int {\frac {d^3\vec{p}}{(2\pi)^3}}
{\frac {\sqrt{p^2+M^2}}{(p^2+M^2)^n}} \;.
\end{equation}
Here, $p=|\vec{p}|$, and ${\cal N}_M[\cdots]$ denotes normal-ordering with
respect to $M$. Following the Refs.~\cite{11,12,13}, one can first calculate the
energy $\int_x<\varphi|{\cal N}_M[{\cal H}_x]|\varphi>$, then take $\varphi_x$
as a constant $\varphi$, and finally, minimize variationally the energy with
respect to ${\cal P}$ as well as $f$ to obtain the GEP. Consequently,
${\cal P}_x=0$, the Fourier component of $f_{xy}$ is $f(p)=\sqrt{p^2 + \Omega^2
(\varphi)}$, and the GEP has the following expression
\begin{eqnarray}
V(\varphi)
&=&{\frac {1}{2}}[I_0(\Omega^2)-I_0(M^2)]
  -{\frac {1}{4}}[\Omega^2I_1(\Omega^2)-M^2I_1(M^2)]  \nonumber \\
  &\;& +{\frac {1}{4}}m^2[I_1(\Omega^2)-I_1(M^2)]
                 + {\frac {3}{4}}\lambda[I_1(\Omega^2)-I_1(M^2)]^2
                 \nonumber \\
  &\;&  +3\lambda[I_{1}(\Omega^2)-I_1(M^2)]\varphi^2
        +{\frac {1}{2}}m^2\varphi^2
   + \lambda\varphi^4    \;
\end{eqnarray}
with the gap $\Omega=\Omega(\varphi)$ satisfying
\begin{equation}
\Omega^2=m^2+6\lambda[I_1(\Omega^2)-I_1(M^2)]+12\lambda\varphi^2 \;.
\end{equation}

For the lower-dimensional cases ($D<3$), the corresponding GEP contains no
divergences \cite{13}, and no further renormalization precedure need to be
considered. However, in the present case, the situation is completely different
and the right hands of Eqs.(6) and (7) are still full of divergences. Hence, we
have to find a further renormalization scheme for making the GEP finite. Next,
let us analyse those divergences appeared in Eqs.(6) and (7).

A straightforward calculation can yield
\begin{equation}
I_1(\Omega^2)-I_1(M^2)=-{\frac {1}{8\pi^2}}(\Omega^2-M^2)
   +{\frac {1}{8\pi^2}}\Omega^2ln{\frac {\Omega^2}{M^2}}-
   {\frac {1}{2}}(\Omega^2-M^2)I_2(M^2)
\end{equation}
and
\begin{eqnarray}
&&{\frac {1}{2}}[I_0(\Omega^2)-I_0(M^2)]
  -{\frac {1}{4}}[\Omega^2I_1(\Omega^2)-M^2I_1(M^2)] \nonumber\\
=&&{\frac {1}{128\pi^2}}(\Omega^4-M^4)-{\frac {1}{64\pi^2}}
\Omega^4ln{\frac {\Omega^2}{M^2}}+{\frac {1}{16}}(\Omega^4-M^4)I_2(M^2) \;,
\end{eqnarray}
where $I_2(M^2)$ is a logarithmic divergent integral. Substituting the last two
equations into Eqs.(6) and (7), one can see that a logarithmic divergence
exists in Eq.(7) and the right hand of Eq.(6) has the terms with logarithmic
divergences squared at most, except for divergent constants (independent of
$\varphi$). It is well known that in the case of lower dimensions, Coleman's
normal-ordering renormalization prescription amounts to the renormalization of
the mass parameter \cite{13,2} (1985), and the relation between the bare and
renormalized coupling parameters is in fact a finite relation \cite{2}(1985).
Thus, generally, for the present case, in order to get rid of the divergences
in Eqs.(6) and (7), perhaps we should consider further a real renormalization
of the coupling parameter and an infinite renormalization of the wave-function.
That is to say, the coupling parameter and the field $\varphi$ should be so
appropriately redefined or infinitely re-scaled that those divergences both in
Eq.(6) and in Eq.(7) can cancel out, respectively, except for divergent
constants in Eq.(6).

From the above analysis, we take
\begin{equation}
\lambda=a_1I^{-1}_2(M^2)+a_2I^{-2}_2(M^2), \ \ \ \ \ 
\varphi^2=bI_2(M^2)\Phi^2
\end{equation}
with $a_1$ and $b$ being some constants to be determined. Note that for $b$,
what is necessary is only to determine it up to a finite rescaling of $\Phi$
\cite{3}. In the ansatz Eq.(10), $a_2$ should be a parameter relevant to
renormalized counterpart of the bare coupling $\lambda$, $\lambda_R$, with a
finite relation, if Eq.(10) can really renormalize the theory. Analysing Eqs.(6)
and (7), one can see that the ansatz Eq.(10) should has contained all
possibilities to remove various divergences both in Eqs.(6) and (7), and other
ansatzes would not either remove those divergences or lead to a new result.
Substituting Eq.(10) into Eq.(6), one can find that those divergences in Eq.(7)
cancel out, and Eq.(6) has only two types of divergent terms : $\Phi^2I_2(M^2)$
and $\Phi^4 I_2(M)$, except for the divergent constant ${\cal D}\equiv
V(\varphi=0)$. Then taking the coeficients of $\Phi^2 I_2(M)$ and $\Phi^4 I_2(M)
$ terms as zero, respectively, leads to
\begin{eqnarray}
\biggl\{
\begin{array}{c}
{\frac {b(m^2+3a_1M^2)}{2(1+3a_1)}}=0 \\
{\frac {a_1b^2}{1+3a_1}}(1-6a_1)=0    
\end{array}
\end{eqnarray}
and consequently the GEP in Eq.(6) also contains no divergences, except for
${\cal D}$. Thus, the solutions for the set of equations Eq.(11) will be
helpful to determine the renormalization schemes.

Obviously, there are three solutions for the set of equations Eq.(11):
\begin{eqnarray}
\begin{array}{lllr}
(i). & \ \ a_1\sim I_2(M^2), & \ \ b=I^{-1}_2(M^2) \  ;& \ \ \ \  \\
(ii). & \ \ a_1=-{\frac {1}{3}}, & \ \  b=I^{-1}_2(M^2), & M^2=m^2 \; ;   \\
(iii).& \ \ a_1={\frac {1}{6}}, & \ \ M^2=-2m^2,& b \ {\rm is \ finite} \; .
\end{array}
\end{eqnarray}
In the strict sense of mathematics, $ii$ in Eq.(12) cannot be regarded as a
solution of Eq.(11), because in the second equation of Eq.(11) it leads to the
infinitesimal $b^2$ dividing by zero \footnote{The author thanks the referee
for commenting upon this point.}. Nevertheless, $ii$ in Eq.(12) can really get
rid of those divergences in Eqs.(6) and (7). In fact, in Eqs.(6) and (7) with
Eq.(10), there is not the denominator $(1+3a_1)$ at all, and the appearance of
this denominator is due to a mathematical deformation of rewriting Eq.(7).
Additionally, although the solution ($i$) can make Eqs.(6) and (7) have no
explicit divergences (of course except for a divergent constant in Eq.(6)), it
means that $\lambda$ is finite and therefore does not produce a viable theory,
for the vacuum could become unstable \cite{2}. In view of this, in the
following, we discuss the other two solutions.

The solution ($ii$) in Eq.(12) implies that it is not necessary to perform the
wave-function renormalization. For the convenience of comparison with
Ref.~\cite{2}, let $a_2={\frac {k}{12\pi^2}}$ with $k={\frac {-4\pi^2}
{\lambda_R}}$. The renomalization scheme is
\begin{eqnarray}
\biggl\{
\begin{array}{l}
\lambda=-{\frac {1}{3}}I^{-1}_2(M^2)+
      {\frac {k}{12\pi^2}}I^{-2}_2(M^2) \\
M^2=m^2
\end{array} 
\ .
\end{eqnarray}
Substituting Eqs.(13), (8) and (9) into Eqs.(6) and (7), we have
\begin{equation}
V(\Phi)={\frac {1}{2}}\Omega^2\Phi^2+{\frac {1}{128\pi^2}}
 [2\Omega^4ln{\frac {\Omega^2}{m^2}}-(3-2k)(\Omega^2-m^2)^2-2m^2(\Omega^2-m^2)]
 +{\cal D}_1
\end{equation}
and
\begin{equation}
\Omega^2ln{\frac {\Omega^2}{m^2}}=(\Omega^2-m^2)(1-k)-16\pi^2\Phi^2
\end{equation}
with ${\cal D}_1$ being a divergent constant. Eqs.(14) and (15) are identical
to Eqs.(5.9) and (5.10) in Ref.~\cite{2}(1985), respectively. Thus, we
reproduce ``precarious $\phi^4$'' \cite{2}. However, as has been seen in the
above, only when $M^2=m^2$, can the divergences be removed, and therefore, if
and only if $M^2=m^2$, ``precarious $\phi^4$'' exists. By the way, because
$M^2$ must be positive, ``precarious $\phi^4$'' corresponds to Eq.(1) with
$m^2>0$. That is to say, if and only if the classical mass in the case of
$m^2>0$ and $\lambda<0$ is taken as the normal-ordering mass, ``precarious
$\phi^4$'' can appear. This also implies that ``precarious $\phi^4$'' arises
from only the symmetrical phase of the classical theory.

Now we are in a position to discuss the solution ($iii$). In this case, we take
$a_2=-{\frac {1}{6}}\lambda_R$ and  $b={\frac {1}{2}}$ for the consistence with
Ref.~\cite{3}. Hence, we have the renormalization scheme
\begin{eqnarray}
\biggl\{
\begin{array}{l}
\lambda={\frac {1}{6}}I^{-1}_2(M^2)-{\frac {1}{6}}\lambda_R I^{-2}_2(M^2) 
    \ \\
\varphi^2={\frac {1}{2}}I_2(M^2)\Phi^2 \  \\
M^2=-2m^2
\end{array} 
\ .
\end{eqnarray}
The third equation in Eq.(16) requires that $m^2$ is negative. Employing this
scheme, Eqs.(6) and (7) can be expressed as
\begin{equation}
V(\Phi)=
({\frac {1}{6}}\lambda_R-{\frac {1}{24\pi^2}})m^2\Phi^2+
    {\frac {1}{36}}\lambda_R\Phi^4
    +{\frac {1}{144\pi^2}}\Phi^4
    (ln{\frac {\Omega^2}{-2m^2}}-{\frac {3}{2}}) +{\cal D}_2
\end{equation}
and 
\begin{equation}
\Omega^2={\frac {2}{3}}\Phi^2
\end{equation}
with ${\cal D}_2$ the divergent constant. Eq.(17) is the GEP obtained by the
scheme Eq.(16). A numerical analysis indicates that so long as $\lambda_R$ is
not too small the SSB remains to exist, whereas if $\lambda_R$ is small the
vacuum enjoys its symmetry \cite{13}.

Owing to $I_2(x)-I_2(y)=-{\frac {1}{4\pi^2}}ln{\frac {x^2}{y^2}}$, we can
introduce a finite characteristic scale $\mu$ with the dimension of mass, which
is related to the renormalized coupling with $\lambda_R=-{\frac {1}{4\pi^2}}
ln{\frac {\mu^2}{M^2}}$. Rewriting $\lambda$ in Eq.(16) in terms of $\mu$ :
\begin{equation}
\lambda={\frac {1}{6I_2(\mu^2)}} \;
\end{equation}
we obtain
\begin{equation}
V(\Phi)={\frac {M^2}{48\pi^2}}(1+ln{\frac {\mu^2}{M^2}})\Phi^2+
        {\frac {1}{144\pi^2}}\Phi^4(ln{\frac {\Omega^2}{\mu^2}}-
        {\frac {3}{2}}) \; .
\end{equation}

It is evident that the expression of Eq.(20) is identical to that of
``autonomous $\phi^4$'' in Ref.~\cite{3} when setting $m^2_0={\frac
{M^2}{24\pi^2}}(1+ln{\frac {\mu^2} {M^2}})$. Here, we have seen that in the
present case, if and only if we make a special choice of normal-ordering mass,
$i.e.$, $M^2=-2m^2$,  the GEP is renormalized and ``autonomous $\phi^4$'' exist.
Contrary to ``precarious $\phi^4$'', ``autonomous $\phi^4$'' corresponds to
Eq.(1) with $m^2<0$, and originates from the asymmetrical phase of the classical
theory. That is to say, only when the classical mass in the case of $m^2<0$ and
$\lambda>0$ is chosen as the normal-ordering mass, ``autonomous $\phi^4$''
appears.

Perhaps one must have noticed that $\lambda$ in Eq.(16) or Eq.(19) is just
Eq.(9) in Ref.~\cite{3} \footnote{Note that $I_2(\mu^2)$ here is equal to
$2I_{-1}(\mu)$ in Ref.~\cite{3}.}. In the same way as Eq.(19) is derived, one also
can rewite $\lambda$ in Eq.(13) and find that $\lambda$ in Eq.(13) is just
Eq.(5.5) in Ref.~\cite{2}(1985). Further, simple calculations of $\Omega
(\varphi=0)$ from Eq.(7) and $\lambda$ in Eqs.(13) and (16) indicate that $M^2
=m^2$ is really consistent with the renormalization condition of the mass in
Ref.~\cite{2}, and $M^2=-2m^2$ consistent with Eq.(10) of Ref.~\cite{3}
\footnote{In the case of $m^2<0$, $I_1(\Omega)$ in $\Omega(\varphi=0)$ should
be taken as $I_1(0)$ to keep consistence with Ref.~\cite{3}}. So the
renormalization scheme Eq.(13) is equivalent to that in Ref.~\cite{2}, and the
renormalization scheme Eq.(16) equivalent to the scheme Eqs.(9),(10) and (11)
in Ref.~\cite{2}. That is to say, the two values of the normal-ordering mass in
the treatment here, $m$ for the case of $m^2>0$ and $\lambda<0$ and
$\sqrt{-2m^2}$ for the case of $m^2<0$ and $\lambda>0$, correspond to the
renormalization point $\varphi=0$ chosen for the case of $\lambda<0$ in
Ref.~\cite{2} and for the case of $\lambda>0$ in Ref.~\cite{3}, respectively.
Because other choices of the normal-ordering mass can not yield a renormalized
and non-trivial theory, the existence of both ``precarious $\phi^4$'' and
``autonomous $\phi^4$'' is crucially dependent upon the normal-ordering mass
(or renormalization points).

In the same way, we have also considered the O(N)-symmetric $\lambda\phi^4$
theory and obtained an analogous conclusion. That is, for the O(N)-symmetric
$\lambda\phi^4$ theory, ``precarious'' and ``autonomous'' $\phi^4$s in
Ref.~\cite{4} exist only at $M^2=m^2$ ($m^2>0$) and at $M^2=-{\frac {2m^2}
{-1+\sqrt{N+3}}}$ ($m^2<0$), respectively. Note that the value $M^2=m^2$ is
just the classical mass with the symmetric vacuum and $M^2=-{\frac {2m^2}
{-1+\sqrt{N+3}}}$ the one with the asymmetric vacuum. Thus, we conclude that
the existence of ``precarious'' and ``autonomous'' $\phi^4$s is uniquely at the
respective particular values of the normal-ordering mass the respective
classical masses. Besides, we want to emphasize that as has been shown in the
derivation of this paper, if any other positive value (not the classical mass)
is chosen as the normal-ordering mass, not only the ``precarious'' and
``autonomous'' $\phi^4$s disappear, but also the massive $\lambda\phi^4$ with
(3+1) dimensions is either trivial or non-renormalizable for any $\lambda$
within the framework of the GWFA. Finally, it should be mentioned that because
normal-ordering mass must be greater than zero, the conclusion in this paper is
not valid for the massless $\lambda\phi^4$ theory.

\acknowledgments
The author is greatly indebted to Professor Guang-jiong Ni for his helpful
discussion and carefully reading the initial manuscript. Also, the author
thanks the referee for his/her helpful comment. This project was supported jointly
by the President Foundation of Shanghai Jiao Tong University and the National
Natural Science Foundation of China with grant No. 19875034.

\end{document}